\documentclass[11pt]{article}

\usepackage[margin=1in]{geometry}
\usepackage{setspace}
\usepackage{caption}
\usepackage{amsmath, amssymb, mathtools}

\usepackage{newtxtext,newtxmath}

\usepackage{booktabs}
\usepackage{graphicx}

\usepackage[numbers]{natbib}

\usepackage[colorlinks=true,
            linkcolor=blue,
            citecolor=blue,
            urlcolor=blue]{hyperref}

\usepackage{authblk}

\makeatletter
\renewcommand\AB@affilsepx{\protect\footnotemark}
\makeatother

\title{On the estimation of inclusion probabilities for weighted analyses of nested case control studies}

\author{
Tomeu López-Nieto-Veitch\thanks{Department of Statistics, University of Bologna, Bologna, Italy. Joint first authors}, \quad
Rossella De Sabbata\thanks{School of Economics, University of Bristol, Bristol, United Kingdom. Joint first authors}, \quad
Ryung Kim\thanks{Department of Epidemiology and Population Health, Albert Einstein College of Medicine, Bronx, NY, USA}, \quad
Sven Ove Samuelsen\thanks{Department of Mathematics, University of Oslo, Oslo, Norway}\thanks{Department of Physical Health and Aging, Norwegian Institute of Public Health, Oslo, Norway}, \quad
Nathalie C. Støer\thanks{Department of Research, Cancer Registry of Norway, Norwegian Institute of Public Health, Oslo, Norway}, \quad
Vivian Viallon\thanks{Nutrition and Metabolism Branch, International Agency for Research on Cancer, Lyon, France. Corresponding author. Email: viallonv@iarc.who.int}
}

\date{\today \thanks{
Acknowledgements: This research was supported by grants from the European Union’s Horizon 2020 research and innovation programme under the Marie Skłodowska-Curie grant agreement (ESSGN 101073237).} 
\thanks{ \textit{IARC Disclaimer}: Where authors are identified as personnel of the International Agency for Research on Cancer/World Health
Organization, the authors alone are responsible for the
views expressed in this article and they do not necessarily represent the decisions, policy, or views of the International Agency for Research on Cancer/World Health
Organization.
}}

\newcommand{\gray}[1]{{\textcolor{gray}{#1}}}

\DeclareMathOperator*{\argmin}{argmin}

\usepackage{pdflscape}
\usepackage{tikz}
\usepackage{tikzsymbols}
\usepackage{relsize}
\usetikzlibrary{arrows, decorations.pathmorphing,shapes, calc}

\tikzset{fontscale/.style = {font=\relsize{#1}}
    }

\tikzstyle{vertex0}=[rectangle,fill=white, draw, minimum size=7pt,inner sep=0pt]
\tikzstyle{vertex1}=[circle,fill=blue, draw,minimum size=7pt,inner sep=0pt]
\tikzstyle{vertex2}=[circle,fill=black, draw,minimum size=7pt,inner sep=0pt]
\tikzstyle{var}=[draw,circle,thick, minimum width=7mm, inner sep = 3pt]
\tikzstyle{varrect}=[draw,rectangle,thick, minimum width=6mm, minimum height = 6mm]
\tikzstyle{varr}=[draw,circle,thick]
\tikzstyle{varf}=[draw=none,fill=none]
\tikzstyle{varCond}=[draw,rectangle,rounded corners=3pt, fill=gray, minimum width=2em,minimum height=2em]
\tikzstyle{varText}=[draw,rectangle,rounded corners=3pt,minimum width=6em,minimum height=2em]

\tikzstyle{edge} = [draw,thick,->]
\tikzstyle{edge2} = [draw,thick,-]
\tikzstyle{edge3} = [draw,dashed,->,>=latex]  
\tikzstyle{edge4} = [draw,dashed,<->,>=latex]  
\tikzstyle{edge5} = [draw, thick,<->,>=latex] 

\tikzstyle{edgedashed} = [draw,thick,->, dashed]
\tikzstyle{edgeblue} = [draw,thick,->, blue]
\tikzstyle{edgered} = [draw,thick,->, red]
\tikzstyle{weight} = [font=\small]

\makeatletter
\newcommand*{\indep}{%
  \mathbin{%
    \mathpalette{\@indep}{}%
  }%
}
\newcommand*{\nindep}{%
  \mathbin{
    \mathpalette{\@indep}{\not}
  }%
}
\newcommand*{\@indep}[2]{%
  \sbox0{$#1\perp\m@th$}
  \sbox2{$#1=$}
  \sbox4{$#1\vcenter{}$}
  \rlap{\copy0}
  \dimen@=\dimexpr\ht2-\ht4-.2pt\relax
  \kern\dimen@
  {#2}%
  \kern\dimen@
  \copy0 
} 
\makeatother

\newcommand{\bM}{{\bf M}}
\newcommand{\bm}{{\bf m}}

\newcommand{\bX}{{\bf X}}

\newcommand{\bz}{{\bf z}}
\newcommand{\bZ}{{\bf Z}}
\newcommand{\bV}{{\bf V}}
\newcommand{\bW}{{\bf W}}

\renewcommand{\P}{\mathbb{P}} 
\newcommand{\1}{\mathbb{I}}
\newcommand{\E}{\mathbb{E}} 
\newcommand{\R}{\mathbb{R}}
\begin{document}

\maketitle

\setstretch{1.25}

\begin{abstract}
\noindent Nested case-control (NCC) studies are a widely adopted design in epidemiology to investigate exposure-disease relationships. This paper examines weighted analyses in NCC studies, focusing on two prominent weighting methods: Kaplan-Meier (KM) weights and Generalized Additive Model (GAM) weights. We consider three target estimands: log-hazard ratios, conditional survival, and associations between exposures. While KM- and GAM-weights are generally robust, we identify specific scenarios where they can lead to biased estimates. We demonstrate that KM-weights can lead to biased estimates when a proportion of the originating cohort is effectively ineligible for NCC selection, particularly with small case proportions or numerous matching factors. Instead, GAM-weights can yield biased results if interactions between matching factors influence disease risk and are not adequately incorporated into weight calculation.
Using Directed Acyclic Graphs (DAGs), we develop a framework to systematically determine which variables should be included in weight calculations. We show that the optimal set of variables depends on the target estimand and the causal relationships between matching factors, exposures, and disease risk. We illustrate our findings with both synthetic and real data from the European Prospective Investigation into Cancer and nutrition (EPIC) study. Additionally, we extend the application of GAM-weights to "untypical" NCC studies, where only a subset of cases are included. Our work provides crucial insights for conducting accurate and robust weighted analyses in NCC studies.

\end{abstract}

\vspace{2ex}
\newpage
\section{Introduction}
Nested case-control (NCC) studies offer an efficient alternative to full cohort studies to investigate the relationship between exposures and disease risk \cite{Thomas1977Addendum, breslow1996, borganetal2018}. In the European Prospective Investigation into Cancer and nutrition (EPIC) study for example, several NCC studies were conducted to assess the relationship between various molecular variables, including metabolomics, and the risk of site-specific cancers \cite{ribolietal2002, breeuretal2022}. NCC studies generally include all participants from an originating cohort who develop the disease of interest during follow-up, referred to as \textit{cases}. For each case, $m\geq 1$ \textit{controls} are randomly selected from the corresponding risk set, defined as the set of participants who have not yet developed the disease at the time the case occurs. Controls and cases are often 
matched on relevant factors, such as sex and age at recruitment, to make them comparable and improve statistical efficiency \cite{langholzclayton1994}. 

The association between exposures and disease risk in NCC studies is usually assessed using conditional logistic regression models, which yield
consistent estimates of the hazard-ratio (HR) if the underlying model is a Cox proportional-hazard model \cite{borganetal1995, samuelsen1997}. Alternatively, weighted analyses have been proposed, in which participants’ weights are derived from their estimated probability of being included in the NCC study  \cite{samuelsen1997}.
Compared to conditional logistic regression models, 
weighted analyses can improve statistical efficiency under the proportional-hazard model  \cite{samuelsen1997, chen2001, kim2013, stoersamuelsen2016} and  provide valid estimates of the HR and related quantities, including survival functions, under more general survival models  \cite{samuelsen1997, MarkKatki2006}. They also allow for the estimation of HRs for the matching factors as well as the analysis of secondary outcomes  \cite{kimkaplan2014}. They could also be used to estimate  associations between exposures, although some concerns have been raised in this context  \cite{stoersamuelsen2016}. 

Several methods have been developed for the estimation of inclusion probabilities and the derivation of the corresponding weights. Kaplan-Meier weights
 (KM-weights) were originally proposed for unmatched settings  \cite{samuelsen1997}, and later 
 extended to matched settings  \cite{kim2013} as well as  ``untypical'' NCC studies, where cases are 
 a random subset of subjects who develop the event of interest  \cite{zhouetal2022, edelmannetal2023}. Additionally, model-based approaches have been proposed
  \cite{samuelsenetal2007, saarelaetal2008, stoersamuelsen2013, stoersamuelsen2016}, such as employing
 generalized additive models (GAMs) 
 to derive GAM-weights. Other approaches include local averaging weights  \cite{chen2001},  though they have been
 shown to be potentially
 unstable in matched settings  \cite{stoersamuelsen2016}.

A few studies have compared the empirical performance of weighted analyses based on different types of weights, such as KM- and GAM-weights. These two types of weights have generally been shown to yield similar results, except in situations involving close matching, 
where KM-weights could lead to more biased estimates compared to GAM-weights  \cite{stoersamuelsen2013}. In the present work, we contribute to the literature by highlighting specific cases where analyses based on either KM-weights or GAM-weights can be severely biased. First, we show that analyses based on KM-weights are generally biased when some participants in the originating cohort are effectively ineligible for selection into the NCC study. This typically occurs when the proportion of cases is small or the number of matching factors is large, as illustrated in Section \ref{sec:EPIC_0}. 

Second, analyses based on GAM-weights can produce biased estimates when interactions between matching factors influence disease risk, unless these interactions are properly accounted for when estimating inclusion probabilities. 

Our observations from these two scenarios led us to more broadly investigate which variables should be considered when computing weights. Using a Directed Acyclic Graph (DAG) to represent the causal relationships among relevant variables, we demonstrate that the decision to include a matching factor in the weight calculation depends on the target estimand and the factor's relationship to both the exposures and the disease risk. 
We illustrate our results on both synthetic and real data from the EPIC study, focusing on three target estimands: log-HRs as measures of the association between exposures and disease risk; the conditional survival function at some given value of the exposures; and parameters of linear regression models as measures of the association between exposures. Finally, another contribution of our work is the extension of GAM-weights to untypical NCC studies. 

The rest of the paper is organized as follows. Section \ref{sec:Methods} introduces the framework of NCC studies,  describes the computation of KM- and GAM-weights for both typical and untypical NCC studies, and introduces the EPIC ENDO study, which serves as a motivating and illustrating example. In Section \ref{sec:Sim}, we use synthetic data to illustrate the aforementioned limitations of KM- and GAM-weights, as well as the possibility to ignore some matching factors in the computation of weights. In Section \ref{sec:EPIC}, the EPIC ENDO study is used to compare weighted analyses based on different types of weights and sets of matching factors used in their computation. We conclude with final remarks in Section \ref{sec:Discussion}. The Appendix provides additional technical details. Additional results from simulation studies illustrating the performance of our proposed GAM-weights for untypical NCC studies, along with further discussion on the use of weighted analyses to estimate associations between exposures (or, more generally, any functional of the distribution of exposures), are provided in the Supplementary Materials. 

\section{Weighted analyses of NCC studies and motivating example} \label{sec:Methods}
\subsection{Weighted analyses of NCC studies}
\subsubsection{NCC studies}
Let $Y$ denote the time to the event of interest, $C$ denote the censoring time and $L$ denote left-truncation time. Each  subject $1\leq j\leq n$ of a cohort of size $n\geq 1$ is followed from $L_j$ to $T_j = \min(Y_j, C_j)$, and we define the event indicator as $D_j = \1(T_j = Y_j)$ with $\1(\cdot)$ the indicator function. Participants of the cohort with $D_j =1$ are those who experience the event of interest during follow-up, and are referred to as events, while those with $D_j=0$ are referred to as nonevents. We focus on matched NCCs and we let $\bM_j$ denote a $q$-dimensional vector of matching factors for subject $j$, for some $q\geq 1$. 

Denote by $K = \sum_{j=1}^n D_j$ the total number of events in the cohort. For each event $1\leq i\leq K$, denote the corresponding risk-set by ${\cal R}_i = \{j: 1\leq j\leq n,\ L_j \leq T_i\leq T_j\}$. We define the effective risk set ${\cal R}^{\bM}_i$, from which the $m$ matched controls for event $i$ can be drawn, as the subset of subjects in ${\cal R}_i$ whose matching factors are close enough to $\bM_i$: 
A common approach is caliper matching, where ${\cal R}^{\bM}_i = {\cal R}_i \cap {\cal M}^{\bM}_i$, with ${\cal M}^{\bM}_i = \{j: 1\leq j\leq n: \lvert \bM_j - \bM_i\rvert \leq \varepsilon\}$ for some tolerance vector $\varepsilon\in\R^q$, and with $\bm \leq \varepsilon$ denoting that $\bm$ is less than or equal to $\varepsilon$ component-wise. For illustration purposes, we also consider nearest neighbor matching, which selects the $m = 1$ subjects in ${\cal R}_i$ whose matching factors are closest to $\bM_i$. Let $S_j$ denote the sampling indicator so that $S_j=1$ if subject $j$ is selected in the NCC and 0 otherwise. Further set $S_{1,j} = S_j D_j$ so that $S_{1,j}=1$ indicates that subject $j$ is selected as a case in the NCC study.

NCC studies represent a specific case in which selection in the study is not completely at random but instead depends on variables $D, T,$ and $\bM$. However, the following conditional ignorability of selection condition holds: $S\indep (\bX, Y) | (D, T, \bM)$. As a result, consistency of estimates obtained from weighted analyses of NCC studies is a consequence of the following general result, whose proof is provided in the Appendix. Denote by $\P(S=1| D, T, \bM)$ the probability that an individual is included in the NCC, conditional on their event status $D$, censoring time $T$ and matching factors $\bM$. Assuming that $\P(S=1| D, T, \bM)>0$ almost surely, we have
\begin{equation}
    \E\bigg[\frac{S}{\P(S=1| D, T, \bM)}\phi(\bX, Y)\bigg] = \E[\phi(\bX, Y)], \label{eq:GenPrincipleWeightedAnalyses}
\end{equation}
for any integrable function $\phi$.
Two common strategies for the computation of weights $1/\P(S=1 | D, T, \bM)$ are the KM- and GAM-weights, which are described in the following two paragraphs. 

\subsubsection{Inverse Probability Weights}


\paragraph{KM-weights}
First consider the setting of typical NCCs where all events are selected as cases. Let $n_i$ denote the size of ${\cal R}^{\bM}_i$. If subject $j$ is a nonevent and belongs to ${\cal R}^{\bM}_i$, her probability of being selected as a matched control for case $i$ is given by $m/(n_i-1)$. Because the random selection of matched controls is performed independently for each index case \cite{samuelsen1997}, the inclusion probability for subject $j$ can then be computed by 
\begin{eqnarray}\label{eq:KM-weights}
p_{KM,j}  =  
\begin{dcases}
1 & {\rm if }\ D_j = 1\\
1 - \prod_{i: j\in {\cal R}^{\bM}_i} \left( 1 -  \frac{m}{n_i - 1}\right)  &{\rm if }\ D_j = 0,
\end{dcases}
\end{eqnarray}
which is the extension to matched settings proposed by Kim (2013) \cite{kim2013}. From this, KM-weights can be defined as $1/p_{KM,j}$.  

KM-weights were extended to the setting of untypical NCCs, where only a fraction $\pi_1\leq 1$ of the events are selected as cases  \cite{zhouetal2022, edelmannetal2023}. For example,  Zhou et al. (2022) \cite{zhouetal2022} introduced Horvitz-Thompson (HT) weights based on the following estimates of inclusion probabilities,
\begin{eqnarray}\label{eq:KM-weights_untypical}
p_{HT,j}  = 
\begin{dcases}
\pi_1 + (1-\pi_1) \left[ 1 - \prod_{i: j\in {\cal R}^{\bM}_i} \left( 1 -  \frac{m}{n_i - 1} D_i S_{1,i}\right)\right] & {\rm if }\ D_j = 1\\
1 - \prod_{i: j\in {\cal R}^{\bM}_i} \left( 1 -  \frac{m}{n_i - 1} D_i S_{1,i}\right)  &{\rm if }\ D_j = 0.
\end{dcases}
\end{eqnarray}
In the special case where $\pi_1=1$, $p_{HT,j}= p_{KM,j}$ and the HT-weights reduce to the KM-weights. For simplicity, we refer to both HT- and KM-weights as KM-weights hereafter. 

\paragraph{GAM-weights}
As an alternative to KM-weights, Støer and Samuelsen (2013, 2016) \cite{stoersamuelsen2013, stoersamuelsen2016}, proposed model-based approaches to estimate inclusion probabilities. Focusing on typical NCCs, the conditional expectation of the sampling indicator $S$ in nonevents can be modelled as a function of variables $T$, $L$ and $\bM$ under generalized linear or additive models with logit-link, using data on nonevents available in the cohort. Specifically, Støer and Samuelsen (2016) \cite{stoersamuelsen2016} suggested to use the following GAM model specification for the conditional probability of inclusion in the NCC: 
$$  \P(S=1 | D=0, L, T, \bM) = \frac{1 }{1 + \exp\{-[\alpha + f(L_j) + g(T_j) + h(\bM_j)]\}}, $$
where $\alpha_0$ is an intercept term, $f$, $g$ are some smooth functions and $h(\bm) = \sum_{q=1}^Q h^{(q)}(m_q)$, for some functions $h^{(1)}, \ldots, h^{(Q)}$ used to model each matching factor. Category matched variables can be included as categorical factors while caliper matched variables can be included as smooth functions. The GAM-weights are then defined as $1/p_{GAM,j}$ with
\begin{eqnarray}\label{eq:GAMweights}
  p_{GAM,j}  =  
\begin{dcases}
1 & {\rm if }\ D_j = 1\\
\frac{1 }{1 + \exp\{-[\widehat{\alpha} + \widehat{f}(L_j) + \widehat{g}(T_j) + \widehat{h}(\bM_j)]\}}  &{\rm if }\ D_j = 0
\end{dcases}  
\end{eqnarray}
Contrary to $p_{KM,j}$, which can be seen as a non-parametric estimate and equals the probability that participant $j$ is effectively selected in the NCC, $p_{GAM,j}$ can be seen as a smoothed estimate of this probability.

For untypical NCCs, to account for potentially different inclusion mechanisms, we propose using separate GAM model specifications for events ($\nu = 1$) and nonevents ($\nu = 0$) of the form
$$  \P(S=1  | D=\nu, L, T, \bM) = \frac{1 }{1 + \exp\{-[\alpha_{\nu} + f_{\nu}(L_j) + g_{\nu}(T_j) + h_{\nu}(\bM_j)]\}}, \quad \nu \in \{0, 1\}$$
Then, we define the GAM-weights for untypical NCCs as $1/\tilde p_{GAM,j}$, with
 \begin{equation}\label{eq:GamWeights_untypical}
 \tilde{p}_{GAM,j}  =  \frac{1 }{1 + \exp\{-[\widehat{\alpha}_{D_j} + \widehat{f}_{D_j}(L_j) + \widehat{g}_{D_j}(T_j) + \widehat{h}_{D_j}(\bM_j)]\}}
 \end{equation}
 which are estimated separately in the sample for events ($D_j = 1$) and nonevents ($D_j = 0$). We provide an empirical assessment of weighted analyses for untypical NCC studies based on these weights in the Supplementary Materials.



\subsection{Motivating example: a nested case-control study of endometrial cancer in EPIC} \label{sec:EPIC_0} 

The European Prospective Investigation into Cancer and Nutrition (EPIC) study is a large prospective study of over 500,000 men and women recruited in 1992–2000 in 10 European countries  \cite{ribolietal2002}. It was originally designed to investigate the relationship between diet and cancer risk. Incident cancer cases were identified through a combination of methods including linkage to health insurance records and cancer registries. Information on relevant socio-demographic, anthropometric and lifestyle variables were collected at recruitment for the entire cohort. Around 386,000 participants further provided a blood sample at recruitment, allowing the 
measurement of molecular data, including clinical biomarkers, GWAS, and metabolomics data, in several cancer-specific case-control studies nested within the EPIC study  \cite{breeuretal2022}. For illustration, we consider a nested-case control study focusing on targeted metabolomics data (AbsoluteIDQ p180 commercial kit; Biocrates Life Science AG, Innsbruck Austria) measured in 853 EPIC participants who developed endometrial cancer and 853 matched controls  \cite{dossusetal2021}. This study, referred to as EPIC-ENDO hereafter, focused on women who did not undergo a hysterectomy, and excluded participants from Greece, Sweden and Denmark. A total number of 141,411 women met these criteria, forming the originating cohort from which the NCC was drawn. Among the 141,411 women of the originating cohort, 873 women developed endometrial cancer during the follow-up, from which 853 cases were randomly selected. For each case, one control was randomly selected using risk-set sampling and matching on recruitment center, menopausal status (pre-/peri-/post- menopausal), age at recruitment ($\pm$ 6 months), date of blood collection ($\pm$ 1 month), time of the day of blood collection ($\pm$ 1h), fasting status ($<$3h/ 3–6h/ $>$6h), and, for premenopausal women, phase of menstrual cycle (follicular/ peri-ovulatory/ luteal) 

\begin{figure}[t]
\includegraphics[scale=0.8]{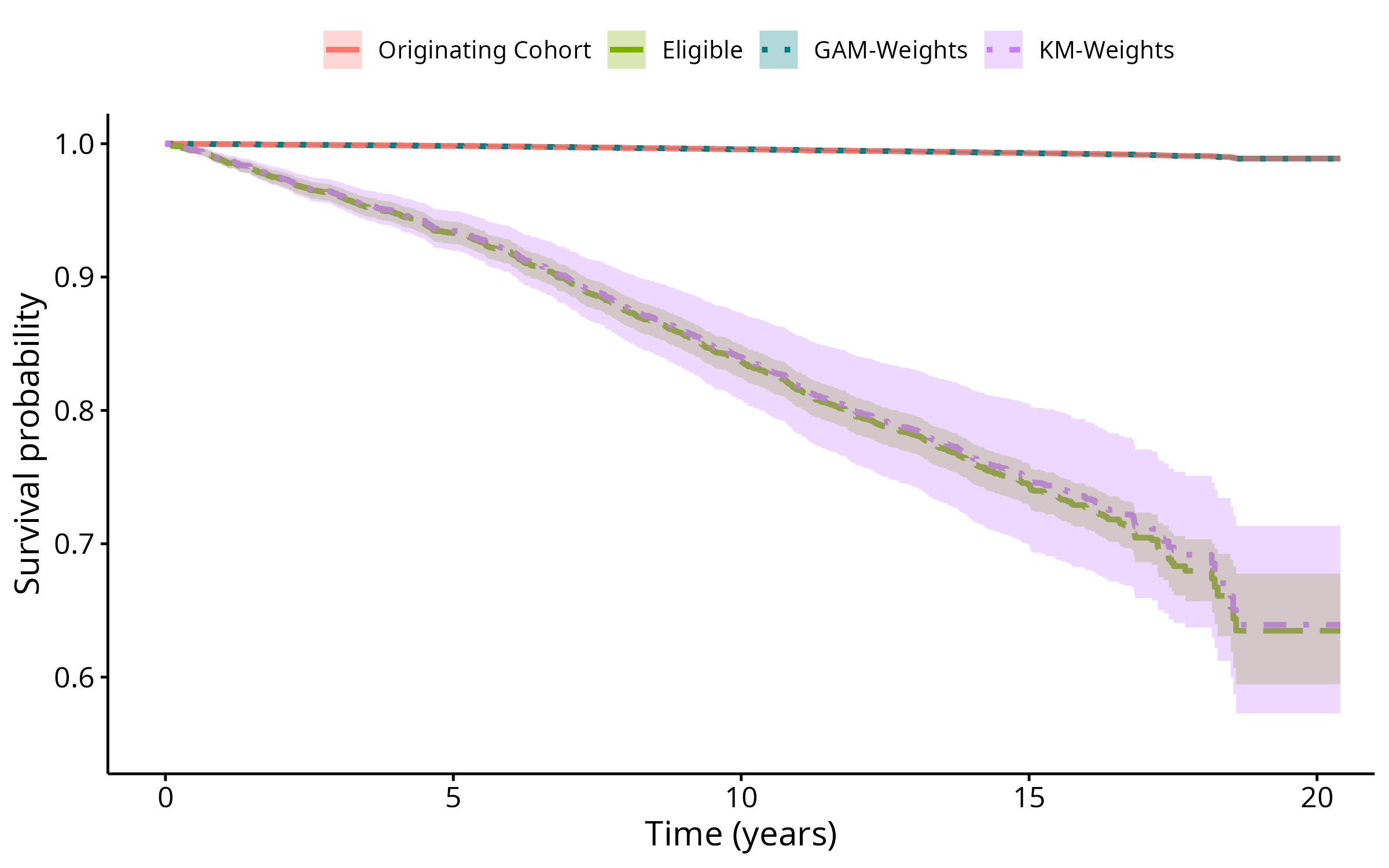} 
\caption{Kaplan-Meier estimates of the absolute survival probability of endometrial cancer, $\P(Y\geq t)$,   produced by the unweighted analysis of the originating cohort of 141,411 EPIC participants (solid line), the unweighted analysis restricted to the 3,432 eligible participants (long-dashed line), and the weighted analyses of the 1,706 participants selected in EPIC-ENDO, using either GAM-weights (dotted line) or KM-weights (dash-dotted line). In this illustration, we considered time on study as the time scale.}\label{fig:Survival_Est_ENDO}
\end{figure}

EPIC-ENDO is an example of an untypical NCC study with $\pi_1 = 853/873 = 0.977$. More importantly, it is a study with small number of events and large number of matching factors, such that a large proportion of participants have a zero probability of being selected. Specifically, 137,979 participants from the originating cohort neither developed endometrial cancer nor belonged to the effective risk sets of any of the 853 cases selected in EPIC-ENDO. Consequently, $\sim 97.6\%$ of the originating cohort was effectively not eligible for selection in EPIC-ENDO.
Figure \ref{fig:Survival_Est_ENDO} shows estimates of the absolute survival probability of endometrial cancer, $\P(Y\geq t)$, obtained in the unweighted analysis of the originating cohort (solid line), in the unweighted analysis of the 3,432 eligible participants (long-dashed line), and in the weighted analysis of the 1,706 participants selected in EPIC-ENDO, using either GAM-weights (dotted line) or KM-weights (dash-dotted line). The solid and dotted curves are indistinguishable, suggesting that the analysis based on GAM-weights is accurate. In contrast, the dash-dotted curve deviates from the solid one noticeably, indicating that the weighted analysis based on KM-weights produces severely biased absolute risk estimates. On the other hand, the dash-dotted and long-dashed curves are very similar, indicating that the weighted analysis based on KM-weights reproduces what is observed in the eligible sub-population. 

We should recall that inverse probability weights used in weighted analyses are intended to create a pseudo-population that reflects the originating cohort.
However, participants who were ineligible for selection do not share all matching characteristics with any of the cases or matched controls. From the perspective of the non-parametric approach underlying Kaplan-Meier (KM) weights, ineligible participants cannot be represented by any of the matched controls. The resulting pseudo-population therefore excludes non-eligible individuals and differs from the originating cohort. Consequently, KM-weighted analyses can only estimate absolute risks representative of the eligible population. This is illustrated by the close similarity between the dash-dotted and long-dashed curves in Figure~\ref{fig:Survival_Est_ENDO}. A simple way to verify this is by computing the sum of the KM-weights over the NCC study, which here equals $~3,420$, similar to the size of the eligible sub-population. By contrast, the sum of the GAM-weights is $~137,504$, close to the size of the originating cohort. From the perspective of the smooth approach used to compute GAM-weights, some participants selected in the NCC can act as representatives of those ineligible for selection, so that the resulting pseudo-population coincides with the originating cohort. This difference between KM- and GAM-weights is exemplified in Supplementary Figure \gray{1} under the simple setting of nearest-neighbor matching: the KM-weights approach yields abrupt jumps in inclusion probability, whereas the GAM-weights approach produces smoothly varying probabilities that capture gradual changes in similarity across individuals.


This illustrative example highlights the importance of the modeling underlying the computation of inverse probability weights. It also raises the following questions, which we address throughout the rest of this article. First, it is easy to realize that when it differs from the originating cohort, the eligible sub-population over-represents cases compared to the originating cohort leading to biased estimation of absolute risk for weighted analyses based on KM-weights. In Section \ref{sec:Sim}, we describe settings where KM-weights also lead to biased assessments of the association between exposures and disease risk and of the associations between exposures, while GAM-weights produce unbiased estimates. A second natural question is whether GAM-weights can lead to biased results in situations where KM-weights do not. We show that when matching factors interact with each other on disease risk, GAM-weights based on Equations \eqref{eq:GAMweights} or \eqref{eq:GamWeights_untypical} generally lead to biased estimates. A third, more general question concerns the variables that should be considered when estimating inclusion probabilities. Using arguments similar to those justifying Equation \eqref{eq:GenPrincipleWeightedAnalyses}, it follows that using weights based on $\P(S=1|D, T, \bM_0)$ leads to consistent estimates for any set $\bM_0$ of variables satisfying $(\bX, Y) \indep S | (D, T, \bM_0)$ and $\P(S=1|D, T, \bM_0) >0 $ almost surely. Specifically, we show that, although selection in the NCC effectively depends on $D, T$ and all matching factors $\bM$, certain matching factors can be excluded from the weight computation under specific conditions.

\section{Illustration on synthetic data}\label{sec:Sim}
We designed a comprehensive simulation study to address the main questions outlined above and to evaluate 
the performance of weighted analyses using different versions of the GAM- and KM-weights. The general design of this simulation study is described in Section \ref{sec:GeneralSettingSim}. In Section \ref{sec:SimKMbiased} we illustrate the limitations of weighted analyses based on KM-weights in the case of nearest-neighbor matching, where some participants of the originating cohort have a null probability of inclusion. In Section \ref{sec:KM_vs_GAM_Inter}, we consider settings where interactions between matching variables influence the event risk, illustrating the 
limitations of analyses based on GAM-weights that ignore such interactions. 
In Section \ref{sec:IgnoreM}, we present simulation results that identify scenarios in which matching variables can be omitted from the computation of GAM- and KM- weights. We further complement these analyses with an exploration of the performance of weighted analyses in untypical NCC designs and their limitations in estimating the association between exposures. These additional settings are presented in the Supplementary materials. 


\subsection{General framework considered in our simulation study}\label{sec:GeneralSettingSim}

\subsubsection{Data Generation}\label{sec:SimDataGen}
\begin{figure}[t]
    \centering
    \begin{tikzpicture}[scale=1, auto,swap]
\node[var] (X1)at(0.5,0){{$X_a$}};
\node[var] (X2)at(1.5,1.25){{$X_b$}};
\node[varrect] (Y)at(3,0){{$Y$}};
\node[var] (T)at(6,0){{$T$}};
\node[varrect] (C)at(4.5,1.25){$C$};
\node[var] (D)at(4.5,-1.25){{$D$}};
\node[var, fill=gray] (S)at(4.5,-3.00){$S$};
\node[var] (V)at(1.75,-2.25){$M$};
\draw[edge, dotted] (V)--(Y);
\draw[edge, dotted] (V)--(X2);
\draw[edge, dotted] (V)--(X1);
\draw[edge] (V)--(S);
\draw[edge, dotted] (X1)--(Y);
\draw[edge] (X1)--(X2);
\draw[edge, dotted] (X2)--(Y);
\draw[edge] (Y)--(D);
\draw[edge] (T)--(D);
\draw[edge] (Y)--(T);
\draw[edge] (C)--(T);
\draw[edge] (D)--(S);
\draw[edge] (T)..controls(6.25,-1.25)..(S);

\end{tikzpicture}
    \caption{Data generating mechanism considered in our simulation study. Solid arrows represent relationships that are consistently present across all the scenarios we consider in our simulation study. Conversely, dotted arrows represent relationships that may be present or not, depending on the values of specific parameters in our model. Right-censored variables are represented within boxes, while non-censored variables are represented within circles. Variable $S$, which represents selection in the NCC study, is shaded. }
    \label{fig:DAGSimul}
\end{figure}
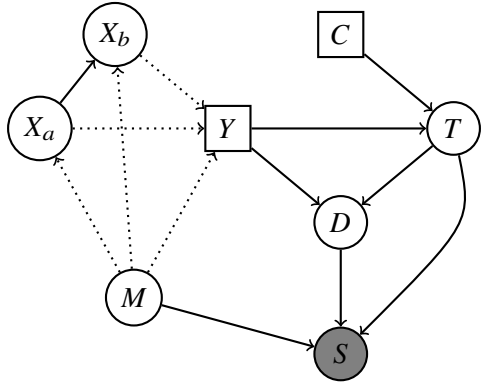

\textbf{Originating Cohort:} We first simulate a cohort of $n=100,000$ participants under the generating model depicted in Figure \ref{fig:DAGSimul}. Specifically, $\textbf{M} = (M_1, M_2)$ represents the vector of matching factors used in the selection of the NCC study, where $M_{1}$ and $M_2$ are centered and scaled versions of a uniform variable ${\cal U}[0,1]$ and a binary variable $\{0,1\}$ with expected value \(\tfrac{1}{2}\), respectively. Variable $M_1$ is used for matching in all scenarios, while $M_{2}$ is also used for matching in Section \ref{sec:KM_vs_GAM_Inter} only. Exposure $X_a$ is modeled as a 3-level categorical variable that mimics a SNP with minor allele frequency (MAF) equal to MAF = $0.25 + \rho_{MX_a}\1(M_1\leq -1)$ in the univariate $M$ case and MAF = $0.25 + \rho_{MX_{a}}\1(M_1\leq 0)\1(M_2=0)$ in the bivariate case. In other words, $X_{a} \in \{0, 1, 2\}$ with probabilities $\P(X_a = 2) = {\rm MAF}^2$ and $\P(X_a = 1) = 2\times{\rm MAF}\times(1-{\rm MAF})$. Exposure $X_b$ is generated under a Gaussian distribution with variance $\sigma^2$ and mean $X_a(\beta + \gamma M_1)$,  with $\beta, \gamma$ and $\sigma^2$ chosen so that the proportion of variance of $X_b$ explained by variables $X_a$ and $M$ is 0.1. After centering and scaling $X_a$ and $X_b$, survival times $Y$ are generated under a Cox proportional hazards model with the hazard rate set to 
\begin{align} \label{eq:surv_times_eq}
   &\lambda(t; X_a, X_b, M_1, M_2) = \nonumber \\
   & \quad \lambda_0(t) \exp(\alpha_a X_a + \alpha_b X_b + \alpha_{M_1} M_1 + \alpha_{M_2} M_2+ \alpha_{M_{1}M_{2}}M_{1} M_{2} + \alpha_{M\cdot X_{a}}M_{1} M_{2} X_a)
\end{align}
for specific values of parameters $\alpha_a, \alpha_b$, $\alpha_{M_1}$, $\alpha_{M_2}$, $\alpha_{M_{1}M_{2}}$ and $\alpha_{M\cdot X_{a}}$ and the baseline hazard rate $\lambda_0(t) = 3t^2/70^3$, corresponding to the hazard rate of a Weibull distribution with shape parameter $k=3$ and scale parameter $\lambda=70$. 
Under this generating mechanism, we can include matching factors with different relationships to the exposure and outcome, allowing us to assess method performance and bias across various scenarios. Depending on the values of $\rho_{MX_a}$, $\gamma_{MX_b}$, $\alpha_{M_1}$, $\alpha_{M_2}$, $\alpha_{M_1M_2}$ and $\alpha_{M\cdot X_a}$, $M_1$ and $M_2$ may influence both the exposure $\mathbf{X}$ and the outcome $Y$ (denoted as $\mathbf{W}$), only the outcome $Y$ (denoted as $\mathbf{V}$), or neither (denoted as $\mathbf{Z}$), as illustrated in Figure \ref{fig:generalDAG}. In Sections \ref{sec:SimKMbiased} and \ref{sec:IgnoreM}, $\alpha_{M_1M_2}$ and $\alpha_{M\cdot X_a}$ are both set to 0, and $M_2$ is not used for matching, so that this variable can be ignored. 

Censoring times $C$ are generated under a ${\cal U}[u_0, u_1]$ distribution for specific values $(u_0, u_1)$. Results are shown for $u_0=20$ and $u_1=50$; other values of these parameters lead to similar results for the estimation of log-hazard ratios and the conditional survival function, and are therefore omitted\footnote{For more details on the influence of $u_0$ and $u_1$ -- and more generally of the censoring rate -- on the estimation of the association between $X_a$ and $X_b$, we refer to Section \gray{4.2} in the Supplementary Materials.}. Finally, observed times and event indicators are defined as $T = \min(Y, C)$ and $D = \1(Y = T)$, respectively. For simplicity, we do not consider delayed entry. The synthetic cohort eventually corresponds of $n = 100,000$ observations of the vector of variables $(M_{1}, M_{2}, X_a, X_b, T, D)$.
\\\\
\textbf{Nested case control study:} NCC studies are selected from the synthetic cohorts as follows. Events ($D=1$) are selected as cases with probability $\pi_1 =1$ for typical NCCs, and probability $\pi_1=0.5$ for untypical NCCs (see Section \gray{3} in the Supplementary Materials). For each selected case $i$, $m=1$ control is randomly selected from the risk-set ${\cal R}_i$, through matching on $M_1$ in Sections \ref{sec:SimKMbiased} and \ref{sec:IgnoreM} and through matching on $(M_1,M_2)$ in Section \ref{sec:KM_vs_GAM_Inter}. Standard caliper matching is considered for  $M_{1}$  in Sections \ref{sec:SimKMbiased}, \ref{sec:KM_vs_GAM_Inter} and \ref{sec:IgnoreM}, using a tolerance $\varepsilon = \sqrt{3}/5$ so that, for any case $i$, the cardinality of ${\cal R}_i\cap\{l: \lvert M_{1,l} - M_{1,i}\rvert \leq \varepsilon\}$ is $20\%$ of that of the risk-set ${\cal R}_i$ on average (since $M_1$ corresponds to a centered and scaled ${\cal U}[0,1]$-variable). In Section \ref{sec:KM_vs_GAM_Inter}, we further consider exact matching for the categorical matching factor $M_{2}$, so that the effective risk set for case $i$ is $\mathcal{R}_i^{M_{1}, M_{2}} = \{ l \in \mathcal{R}_i : \lvert M_{1,l} - M_{1,i} \rvert \leq \varepsilon \} \cap \{ l \in \mathcal{R}_i : M_{2,l} = M_{2,i} \}$. To illustrate the performance of weighted analyses when inclusion probabilities are null for some subjects, we also consider a nearest-neighbor (NN) matching scheme in Section \ref{sec:SimKMbiased}, whereby the matched control is selected from the effective risk-set $\tilde {\cal S}_i = \argmin_{l\in R_i} \lvert M_{1,l} - M_{1,i}\rvert$. Because $M_{1}$ is a continuous variable, the effective risk-set ${\cal S}_i$ reduces to one single subject under the nearest-neighbor (NN) matching scheme and, 
for typical NCCs ($\pi_1=1$), inclusion probabilities are null for all subjects who are not selected in the NCC.


 
\subsubsection{Weights for the weighted analyses}

We consider weighted analyses of NCC studies using KM-weights and GAM-weights, as defined in Equations \eqref{eq:KM-weights} and \eqref{eq:GAMweights} for typical designs (\( \pi_1 = 1 \)), and in Equations \eqref{eq:KM-weights_untypical} and \eqref{eq:GamWeights_untypical} for untypical designs (\( \pi_1 < 1 \)). In Section \ref{sec:KM_vs_GAM_Inter}, we additionally consider a version of GAM-weights that includes the interaction between matching variables, referred to as \textit{``GAM-weights with InterM''}\footnote{The interaction between the continuous matching variable \( M_1 \) and the binary matching variable \( M_2 \), is modeled as a smooth term of $M_1$ that varies by the level of \( M_2 \).}.
 
 Furthermore, in Section \ref{sec:IgnoreM}, we also consider both GAM-weights and KM-weights that ignore the matching factors \( \textbf{M} \). The former are derived from estimates of \( P(S = 1 \mid D, T) \) instead of \( P(S = 1 \mid D, T, M) \), and are referred to as ``\textit{GAM-weights w/o M}''. The latter correspond to versions of KM-weights where \( \mathcal{R}^{\bM}_i \) is replaced by \( \mathcal{R}_i \) in Equations \eqref{eq:KM-weights} and \eqref{eq:KM-weights_untypical}, and are labeled ``\textit{KM-weights w/o M}'' in our figures.


\subsubsection{Evaluation criteria}
We focus on three target estimands: $(i)$, the log-HR $\alpha_b$ of $X_b$, $(ii)$, the conditional survival probability $\P(Y\geq u_1 | X_a = 0)$ at the end of follow-up $u_1$, and $(iii)$, the parameter $\theta$ in the linear regression of $X_b$ on $X_a$ as a measure of the marginal association between $X_a$ and $X_b$. In weighted analyses, these estimands are respectively estimated $(i)$, using the Cox proportional hazard model adjusted for $X_a$ and $M$\footnote{For comparison, we also estimate the log-HR coefficient for $X_a$, $\alpha_b$, using a conditional logistic regression model $X_a$, which is shown to be unbiased also under closed matching (e.g. \cite{stoersamuelsen2013})}., $(ii)$ using the Kaplan-Meier estimator computed in the subgroup of subjects with $X_a = 0$, and $(iii)$, using the (unadjusted) linear regression model of $X_b$ on $X_a$. 


For each target estimand, we treat the estimate from the unweighted full cohort as the gold standard. We evaluate performance by computing the difference between this reference estimate and the corresponding estimate from the nested case-control (NCC) sample.


In each simulation setting, corresponding to a specific combination of parameters introduced in Section \ref{sec:SimDataGen}, we generate 100 cohort studies. From each cohort, we draw a single NCC sample and compute both the full cohort and weighted NCC estimates. We then assess performance based on the observed distribution of the differences between the two sets of estimates across the 100 simulations. We additionally report the means, standard deviations and statistical differences relative to the full cohort estimates in Supplementary material \gray{2}.


\subsection{Nearest-neighbor versus caliper matching} \label{sec:SimKMbiased}

We start with the illustration of the estimation accuracy of weighted analyses in the case of nearest-neighbor matching to illustrate possible limitations of KM-weights when inclusion probabilities are null for some subjects of the originating cohort. The case of caliper matching is also considered for comparison. We consider the standard setting of typical NCCs ($\pi_1 =1$) and generate synthetic cohorts with a single matching factor $M_1$ and parameters $\alpha_{M_1M_2}$ and $\alpha_{M\cdot X_a}$ both set to 0. Here, we present results for the choice $\rho_{MX_a} = 0.2, \gamma_{MX_b}= 0$ and $\alpha_{M_1} = \alpha_a= \alpha_b =\log(2)$. Other values of these parameters lead to similar results, which are therefore omitted. 

\begin{figure}[t] 
    \centering
    \includegraphics[scale=0.2]{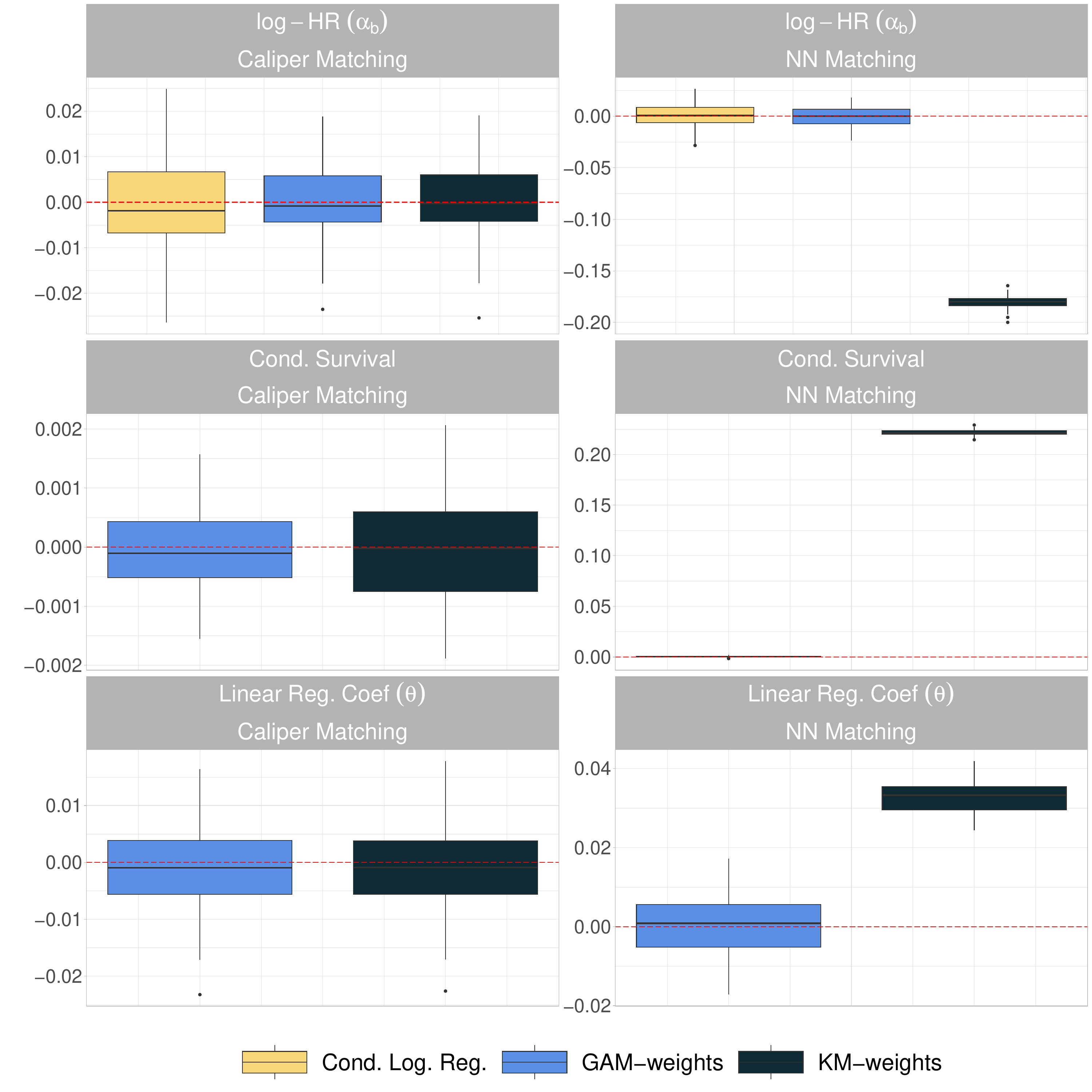}
    \caption{Results of the simulation study illustrating the bias of analyses based on KM-weights in settings where inclusion probabilities are null for some subjects of the cohort. Each boxplot represents the observed distribution of the differences between the estimate produced by one particular type of analysis of the NCC study and the estimate obtained on the full cohort over 100 replicates.}
    \label{fig:NNMatching}
\end{figure}

Figure \ref{fig:NNMatching} presents the results for the estimation of the three estimands of interest. Weighted analyses using standard GAM-weights produce unbiased estimates for all estimands under both types of matching scenarios. Conversely, although they produce unbiased estimates in the case of caliper matching, analyses based on KM-weights produce biased estimates for all estimands in the case of NN-matching. These results are consistent with what was observed in Section \ref{sec:EPIC_0} and confirm that KM-weights are generally not appropriate when inclusion probabilities are null for some subjects of the cohort. The bias in the estimation of $\P(Y\geq u_1| X_a = 0)$ when using KM-weights with NN-matching was here again expected since all the cases are selected and only subjects who do not develop the disease of interest can have a null inclusion probability. Therefore, the sub-population of eligible subjects comprise all events but only a subset of the non-events and $\P(Y\geq u_1| X_a = 0)$ is larger in the selectable population compared to the full cohort. Moreover, in the NN-matching setting we consider here, KM-weights simply equal 1 for all individuals selected in the NCC so that the weighted analysis of the NCC based on KM-weights reduces to the unweighted analysis of the NCC study, and the observed bias in the estimation of each target estimand is a direct consequence of collider bias.

\subsection{Interaction Between Matching Factors and Disease Risk}\label{sec:KM_vs_GAM_Inter}
We now consider two matching factors \( M_1 \) and \( M_2 \) that interact with each other $(M_{1}M_{2})$ and with exposure $X_a$ $(M_{1}M_{2}X_{a})$ on the disease risk, to illustrate potential limitations of GAM-weights in the presence of interactions that affect disease risk. As in the previous section, we present results for the setting $\{ \rho_{MX_a}=0.2, \gamma_{MX_b}= 0, \alpha_a= \log (2), \alpha_b =\log(2) , \alpha_{M_1} = \alpha_{M_2} =0\}$. Other values of these parameters lead to similar results and are therefore omitted. To comprehensively examine the impact of interactions, we consider all combinations of $\alpha_{M1M2}  \in \{0, \log(2)\}$ and $\alpha_{M\cdot X_{a}}  \in \{0, \log(2)\}$.

\begin{figure}[h] 
    \centering
    \includegraphics[width=\linewidth]{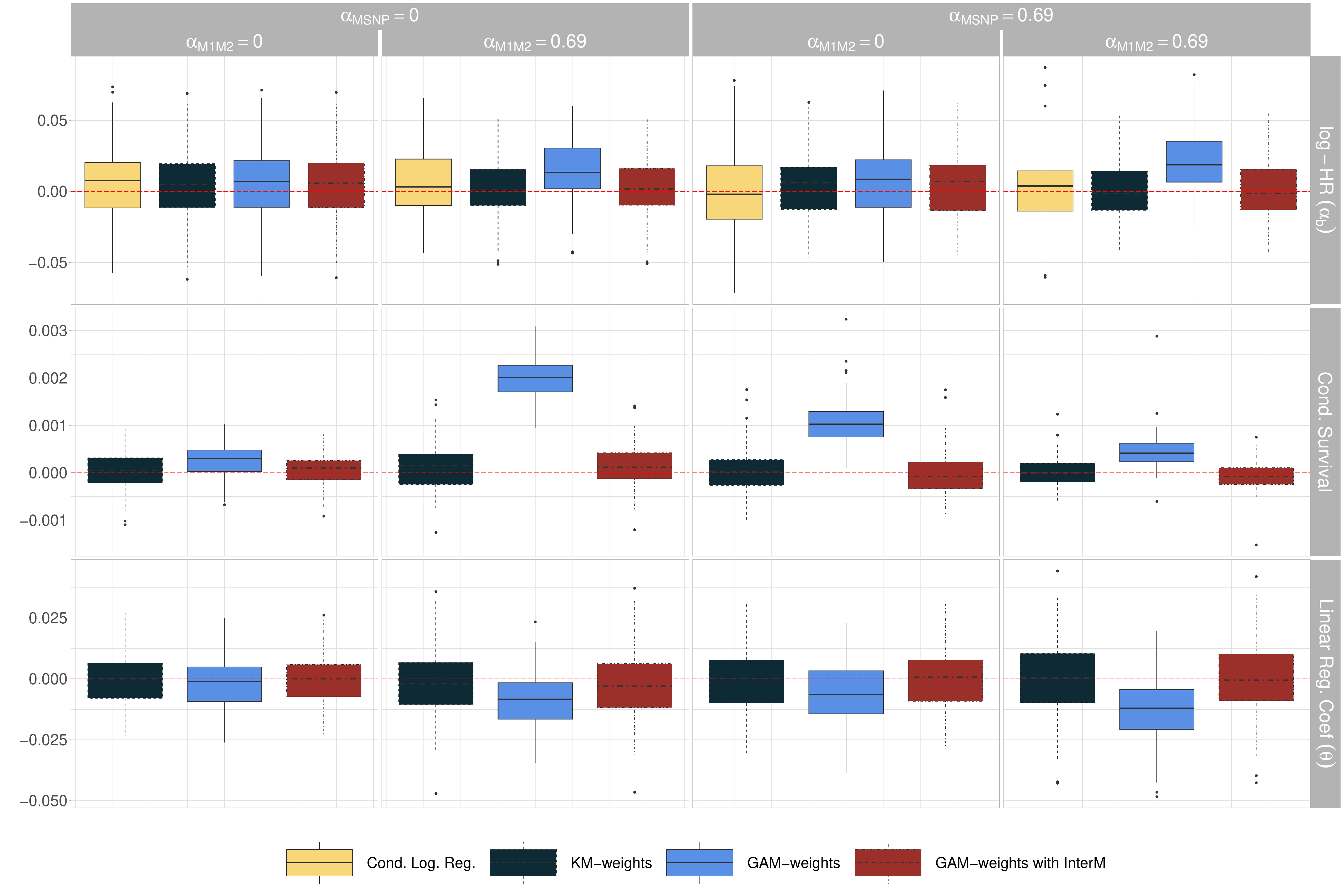}
    \caption{Results of the simulation study incorporating interactions between matching factors on disease risk. Each boxplot shows the observed distribution of the differences between estimates from various NCC analyses and the corresponding full cohort estimate, based on 100 simulation replicates.}
    \label{fig:Inter_KM_vs_GAM_SNP0}
\end{figure}

In Figure \ref{fig:Inter_KM_vs_GAM_SNP0}, we show that 
standard GAM-weights produce substantially biased estimates across all three estimands of interest, whenever the interaction between the matching factors influences survival time. Notably, standard GAM-weights estimates of conditional survival are severely biased even when the interaction between the matching factors does not influence survival time, provided there is an interaction between the matching factors and the SNP ($\alpha_{M\cdot X_{a}} = 0.69$). In contrast, KM-weights as well as GAM-weights including interactions produce unbiased estimates in all settings.

\subsection{When Matching Factors May Be Excluded from Weight Computation}\label{sec:IgnoreM}
\begin{figure}[h] 
    \centering
    \includegraphics[width=\linewidth]{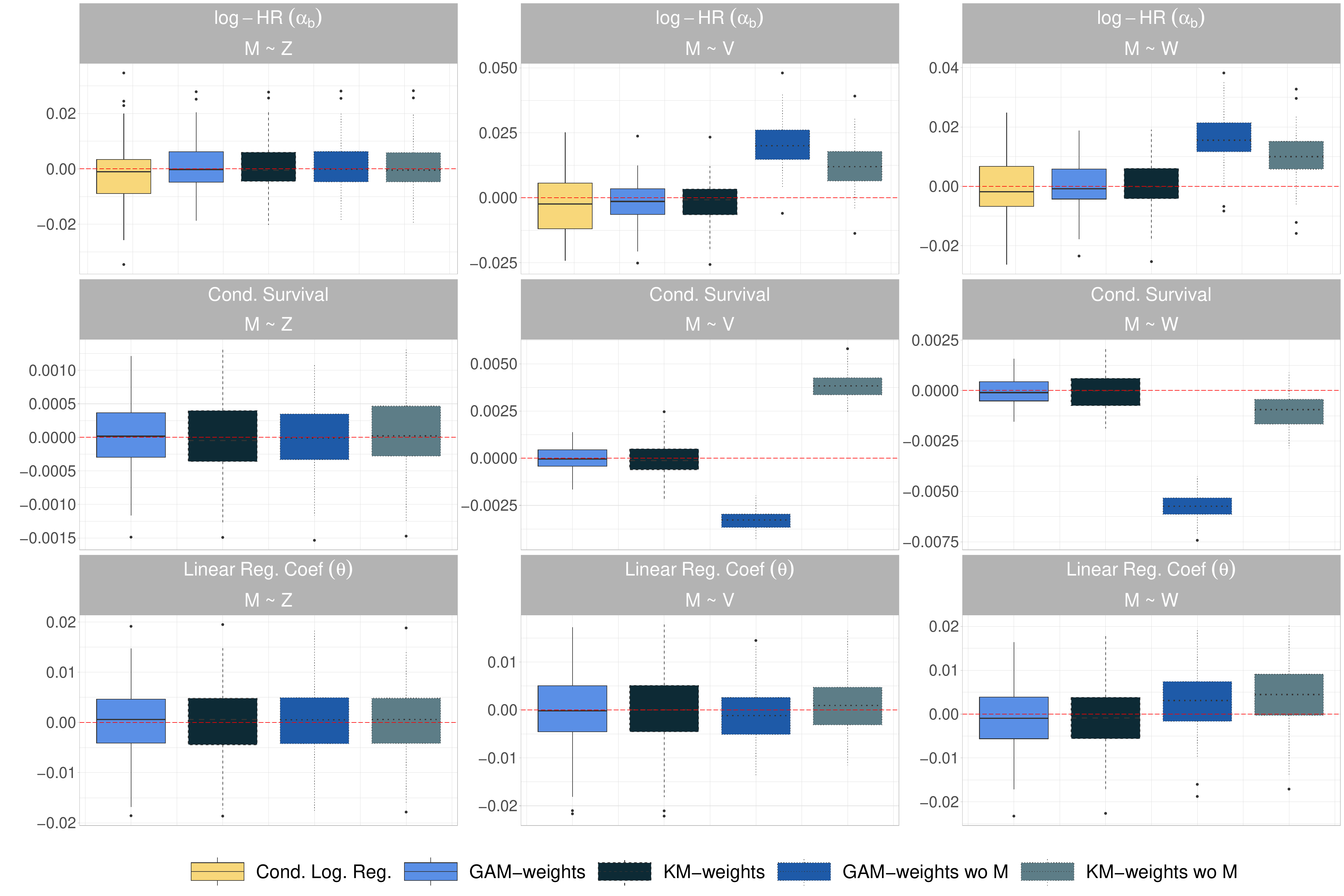}
    \caption{Results of the simulation study illustrating that some matching factors can be ignored in the computations of KM- and GAM-weights. Each boxplot represents the observed distribution of the differences between the estimate produced by one particular type of analysis of the NCC study and the estimate obtained on the full cohort over 100 replicates.}
    \label{fig:IgnoreM}
\end{figure}

We now turn our attention to situations where matching factors can be ignored in the computation of weights. We again consider the standard setting of typical NCCs ($\pi_1 =1$) and generate synthetic cohorts with a single matching factor $\textbf{M}=M_1$ and parameters $\alpha_{M_1M_2}$ and $\alpha_{M\cdot X_a}$ both set to 0. We consider three settings, namely $\{\rho_{MX_a}=0.2, \alpha_{M_1}=\log(2)\}$, $\{\rho_{MX_a}=0,  \alpha_{M_1}=\log(2)\}$, and $\{\rho_{MX_a}=0,  \alpha_{M_1}=0\}$, which corresponds to settings where $\textbf{M}$ behaves as $\bW$ (i.e., influences both $\bX$ and $Y$), $\bV$ (i.e., influences $Y$ only) and $\bZ$ (i.e., influences neither $\bX$ nor $Y$) in Figure \ref{fig:generalDAG}, respectively. Results are presented on Figure \ref{fig:IgnoreM} for the particular choice $\gamma_{MX_b}= 0$, and $\alpha_a= \alpha_b =\log(2)$. Other values of these parameters lead to similar results, which are therefore omitted. 

Results are consistent with our formal derivations presented in \ref{sec:AppendixVariablesToConsider}. Focusing on GAM-weights first, weighted analyses based on standard GAM-weights (that is, with $\textbf{M}$ accounted for in their computation), lead to unbiased estimates for all estimands in all considered settings. Ignoring $\textbf{M}$ in the computation of GAM-weights consistently lead to estimates that are biased when $\textbf{M} \sim \bW$ and unbiased when $\textbf{M}\sim \bZ$, for all estimands. When $\textbf{M} \sim \bV$, estimates of the log-HR and conditional survival function are biased, while those of $\theta$, the association between $X_a$ and $X_b$, are unbiased and with smaller empirical variance compared to those derived when using standard GAM-weights. 
The same patterns are observed for analyses based on KM-weights: ignoring $\textbf{M}$ leads to estimates that are biased when $\textbf{M}\sim \bW$, unbiased when $\bM \sim \bZ$, and biased for the log-HR and conditional survival function but unbiased for $\theta$ when $\textbf{M}\sim \bV$. When $\textbf{M}\sim \bV$, estimates of $\theta$ obtained with KM-weights ignoring $\textbf{M}$ have smaller empirical variance compared to those obtained with standard KM-weights.

\newpage
\section{Illustration on the EPIC ENDO study}\label{sec:EPIC}

We now present additional results from weighted analyses of the EPIC-ENDO study. First, we propose that the matching factors used in EPIC can be grouped in two categories: $(i)$ factors $\bW$, such as recruitment center and menopausal status, which may influence both the metabolite levels $\bX$ and the outcome $Y$, and $(ii)$ ``technical'' factors $\bZ$, such as fasting status, which affect only the measurement process and are assumed to influence neither $\bX$ nor $Y$. 

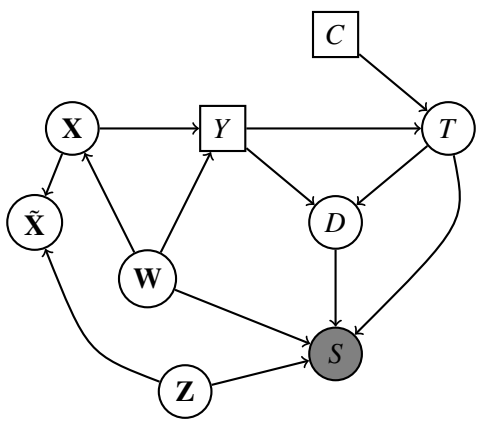
\begin{figure}[t]
    \centering
    \begin{tikzpicture}[scale=1, auto,swap]
\node[var] (X)at(1,0){$\bX$};
\node[var ] (Xm)at(0.5,-1.25){$\tilde\bX$};
\node[varrect] (Y)at(3,0){{$Y$}};
\node[var] (T)at(6,0){{$T$}};
\node[varrect] (C)at(4.5,1.25){$C$};
\node[var] (D)at(4.5,-1.25){{$D$}};
\node[var, fill=gray] (S)at(4.5,-3.00){$S$};
\node[var] (V)at(2.5,-3.50){$\bZ$};
\node[var] (W)at(2,-2){$\bW$};
\draw[edge] (W)--(Y);
\draw[edge] (W)--(S);
\draw[edge] (W)--(X);
\draw[edge] (V)--(S);
\draw[edge] (V)..controls(1.2, -3)..(Xm);
\draw[edge] (X)--(Y);
\draw[edge] (X)--(Xm);
\draw[edge] (Y)--(D);
\draw[edge] (T)--(D);
\draw[edge] (Y)--(T);
\draw[edge] (C)--(T);
\draw[edge] (D)--(S);
\draw[edge] (T)..controls(6.25,-1.25)..(S);

\end{tikzpicture}
    \caption{Schematic representation of the data generating mechanisms in the EPIC cohort. Variables in $\bX$ correspond to the true "underlying" blood levels of metabolites, which are possibly causes of the outcome $Y$. Conversely, variables in $\tilde \bX$ correspond to the measured blood levels of metabolites. These measurements are influenced by "technical variables", such as fasting status and time at blood collection. But, these technical variables do not influence the true "underlying" blood levels $\bX$. Matching factors used in EPIC can be grouped in two categories: $\bW$, such as recruitment center and menopausal status, which are expected to influence both $\bX$ and $Y$, and the purely "technical" factors $\bZ$, which influence neither $\bX$ nor the outcome $Y$. Right-censored variables are represented within boxes, while non-censored variables are represented within circles. Variable $S$, which represents selection in the NCC study, is shaded. }
    \label{fig:DAGEPIC}
\end{figure}

Figure~\ref{fig:DAGEPIC} illustrates the DAG corresponding to this structure in EPIC-ENDO. The variables $\bX$ represent the ``underlying'' or biologically relevant blood levels of metabolites, which are potential causal determinants of the outcome $Y$. In contrast, the observed variables $\tilde\bX$ denote the actual measured metabolite levels, which may be influenced by the technical factors in $\bZ$. Since chronic disease risk is more plausibly linked to sustained metabolic profiles than to short-term fluctuations, $\tilde \bX$ is unlikely to be a cause of $Y$, while $\bX$ may be. Then, we argue that the variables in $\bZ$ do not influence the underlying exposures $\bX$, and cannot plausibly affect the outcome $Y$. For example, while fasting status may transiently alter the measured concentration of metabolites—e.g., leading to higher levels in non-fasting individuals—it is unlikely to influence the long-term average or biologically meaningful blood levels of metabolites $\bX$.

This leads us to consider four types of weights: KM-weights and GAM-weights, as defined in equations \eqref{eq:KM-weights_untypical} and \eqref{eq:GamWeights_untypical}, respectively, as well as variants that exclude technical matching factors, specifically, time and day of blood collection, fasting status, and menstrual cycle phase. In the figures and the text below, these variants are labeled as KM- and GAM-weights wo Technical $M$. Because no interaction between the matching factor and cancer risk is statistically significant in the originating cohort, we do not include interaction terms in the computation of GAM-weights.

Figure \gray{6} in the supplementary material displays the empirical distributions and pairwise correlations of the weights among EPIC-ENDO controls. The weights differ substantially in magnitude and show low correlation, with the exception of the two GAM-based versions, which are nearly identical. As expected, KM-weights are generally smaller than their counterparts that exclude technical factors. GAM-weights are the largest overall, with some participants receiving values exceeding 1000.

We then evaluate results from weighted analyses using the different types of weights, focusing on the following three target estimands: the (unconditional) absolute risk of endometrial cancer, the log-hazard ratio (log-HR) measuring the association between metabolites and endometrial cancer (adjusted for relevant covariates), and associations between exposures. Figure \gray{7} in the supplementary material shows the estimates of the absolute survival probability for endometrial cancer. As already noted in Section \ref{sec:EPIC_0}, estimates based on KM-weights are heavily biased, whereas those derived using GAM-weights perform well. Removing technical matching factors improves the KM-weight estimates, though some bias remains. In contrast, excluding these factors has virtually no effect on GAM-weight estimates.

\begin{figure}[h] 
    \centering
    \includegraphics[trim={7.35cm 0 0 0},clip, scale=0.5]{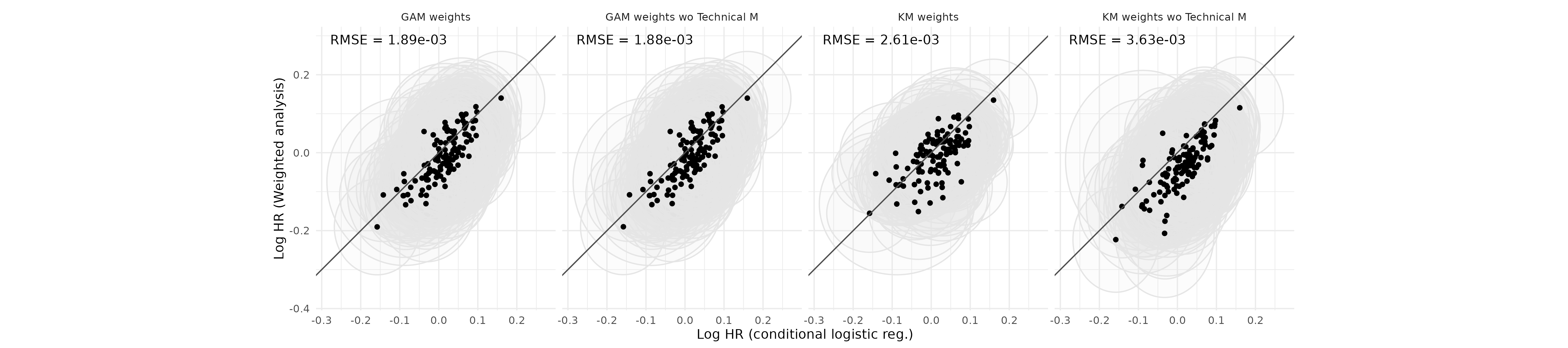}
    \caption{Log-HR estimates produced by weighted analyses across the 117 metabolites measured in EPIC ENDO; comparison with the results from the conditional logistic regression model. For each metabolite, the surrounding ellipse informs on the precision of the estimation. Specifically, the length of the semi-axes corresponds to 1.96sd, with "sd" the standard deviation of the estimate from the conditional logistic regression ($y$-axis) and the weighted analysis ($x$-axis), respectively.}
    \label{fig:EPIC_HR}
\end{figure}

Next, Figure \ref{fig:EPIC_HR} compares the log-HR estimates obtained from conditional logistic regression and from the weighted analyses. The conditional logistic regression models are adjusted for BMI and all caliper matching factors, while the weighted Cox proportional hazards models further include the categorical matching factors. Sensitivity analyses adjusting the conditional logistic regression models for BMI only, or for BMI along with various subsets of the matching factors, yielded similar results and are therefore omitted. Taking the conditional logistic regression estimates as the gold standard, weighted analyses using either version of the GAM-weights produce highly similar estimates and perform best in terms of root mean squared error (RMSE), defined as the sum over the 117 metabolites of the squared differences between the log-HR estimates from the two methods. In contrast, the RMSE increases by approximately 33\% when using KM-weights, and nearly doubles when using KM-weights wo Technical $M$. This indicates that, unlike what was observed for absolute risk estimation, omitting technical matching factors in the KM-weights computation degrades the accuracy of log-HR estimation in EPIC-ENDO.

Lastly, we assess the correlations among exposures. First, we focus on 11 exposures available in the originating cohort: age at inclusion, total energy intake, BMI, the Mediterranean Diet score, habitual alcohol intake, an ordinal smoking score (capturing both habits and intensity), and four physical activity scores—occupational, recreational, vigorous, and total. The corresponding 55 pairwise correlations computed in the originating cohort are considered as reference values, against which we compare estimates derived from the EPIC-ENDO NCC study using several approaches: the unweighted analysis of the full NCC study, the unweighted analysis of the controls only, and the weighted analyses using the four types of weights. For each method, Table \ref{tab:EPIC_Corr_StdExpo} presents the bias (mean difference from the reference correlations) and residual sum of squares (RSS). The unweighted analysis of the full NCC study performs best with respect to both bias and RSS, followed by the unweighted analysis of the controls. This suggests that selection bias has minimal impact. Among the weighted approaches, GAM-weights yield slightly lower bias than KM-weights, but perform worse in terms of RSS. Notably, the RSS for GAM-weights wo Technical $M$ is three times larger than that of the unweighted full NCC analysis.

This finding echoes concerns previously raised in the literature about the use of inverse-probability weighting to infer the distribution of exposures  \cite{stoersamuelsen2016}. In particular, extreme weights can be assigned to controls with short follow-up and can strongly inflate variance.\footnote{Such weights have less impact on log-HR or absolute risk estimation, precisely because the corresponding individuals contribute little follow-up time.} To address this issue, we also consider a thresholded version of the GAM-weights, where weights exceeding 100 are capped at 100. This thresholded version produces bias similar to that of the original GAM-weights but substantially reduces the RSS, bringing it down to the level observed in the unweighted analysis of the controls. Additional results from a comprehensive simulation study are presented in the Supplementary Material. These confirm that thresholded GAM-weights generally yield more stable (i.e., less variable) estimates, although at the cost of a possible increase in bias. The Supplementary Material also includes an analysis of the associations between BMI and the 117 metabolites in the EPIC-ENDO study. Specifically,  Figure \gray{8} in the Supplementary Material presents a heatmap of these correlations, which especially highlights the close similarity between estimates from the unweighted analysis of the controls and those from the weighted analysis using thresholded GAM-weights.

\setlength\tabcolsep{1.8pt}
    \begin{table}[]
    \centering
    \begin{tabular}{|l ccccccc |}
      \hline
      Criterion & Unweight. & Unweight. Ctrls & KM- & KM- wo Tech. M & GAM- & GAM- wo Tech. M & GAM- Tresh.  \\
      \hline
      Bias & 0.009 & 0.015 & 0.024 & 0.018 & 0.016 & 0.015 & 0.015  \\
      RSSE & 0.107 & 0.130 & 0.236 & 0.164 & 0.288 & 0.318 & 0.139  \\
      \hline
    \end{tabular}
    \caption{Bias and residual sum of squared errors (RSSE) for the estimation of pairwise correlations among 11 exposures in EPIC-ENDO.}
    \label{tab:EPIC_Corr_StdExpo}
\end{table}

\section{Conclusion} \label{sec:Discussion}

In this article, we illustrate the performance of weighted analyses of nested case control studies, depending on the type of weights (KM- or GAM-weights) and the variables considered in their computations. Focusing on three target estimands, we show that some bias can arise in certain situations. 

First, our results suggest that caution is required when subjects from the originating cohort can effectively not be selected in the NCC. In such settings, which can occur under specific matching scenarios or when the number of matching factors is large and the number of cases is low, KM-weights can lead to biased estimates, especially for (conditional) absolute risk estimates. GAM-weights, which are based on smoother estimates of inclusion probabilities, can lead to more accurate results in such settings. 

Second, GAM-weights also suffer from limitations in specific settings, especially when matching factors may interact with each other and/or with other exposures on disease risk, unless the computation of inclusion probability properly account for these interactions. More generally, we show that the choice of which variables to consider, and how to model them (in particular, with or without interactions) when computing weights, depends not only on how they affect selection in the NCC study, but also on how they relate to the exposures of interest and disease risk. 

In the absence of established criteria for selecting the best types of weights, we recommend comparing the results of weighted analyses using different weighting schemes (such as KM-weights, GAM-weights, GAM-weights with interactions). Where possible, results of simple analyses can also be compared with those from the corresponding analyses conducted on the full cohort.

Finally, our results confirm previous concerns about the limitations of weighted analyses of NCC studies, particularly when focusing on functionals of the distribution of exposures, such as correlations between exposures. We observe that variance can be substantial, but certain weight variants, in particular thresholded versions and GAM-weights that ignore the variables $T$ and $D$, may offer improved performance. Further work is needed to fully evaluate the performance of different weighting schemes in this context.

\bibliographystyle{SageV}
\bibliography{references.bib}

\newpage
\appendix

\section{Variables to consider in the computation of weights to ensure consistency of weighted analyses of NCC studies}\label{sec:AppendixVariablesToConsider}
In this Appendix, we first introduce a DAG describing the causal relationships among the main variables involved in NCC studies (Figure \ref{fig:generalDAG}). We then prove the general result stated in Equation \eqref{eq:GenPrincipleWeightedAnalyses}. Finally, we present simple results establishing that certain matching factors can be ignored in specific settings, depending on the target estimand. 

\subsection{DAG representing the underlying data generating mechanism}
\begin{figure}[ht]
    \centering
    \begin{tikzpicture}[scale=1, auto,swap]
\node[var] (X)at(0,0){{$\bX$}};
\node[var] (W)at(1.1,-1.75){$\bW$};
\node[varrect] (Y)at(3,0){{$Y$}};
\node[var] (T)at(6,0){{$T$}};
\node[varrect] (C)at(4.5,1.25){$C$};
\node[var] (D)at(4.5,-1.25){{$D$}};
\node[var, fill=gray] (S)at(4.5,-3.5){$S$};
\node[var] (Z)at(2.45,-4.5){$\bZ$};
\node[var] (V)at(1.75,-3.25){$\bV$};

\draw[edge] (W)--(X);
\draw[edge] (W)..controls(2.25,-1.5)..(Y);
\draw[edge] (W)..controls(2.25,-2.75)..(S);
\draw[edge] (V)--(Y);
\draw[edge] (V)--(S);
\draw[edge] (Z)--(S);
\draw[edge] (X)--(Y);
\draw[edge] (Y)--(D);
\draw[edge] (T)--(D);
\draw[edge] (Y)--(T);
\draw[edge] (C)--(T);
\draw[edge] (D)--(S);
\draw[edge] (T)..controls(6.25,-1.25)..(S);

\end{tikzpicture}
    \caption{Typical DAGs in NCC studies, where matching factors can be categorized in 3 groups: factors $\bW$ that may influence both $\bX$ and $Y$, factors $\bV$ that may influence $Y$ only, and factors $\bZ$ that influence neither $\bX$ nor $Y$. Right-censored variables are represented within boxes, while non-censored variables are represented within circles. Variable $S$, which represents selection in the NCC study, is shaded. }
    \label{fig:generalDAG}
\end{figure}
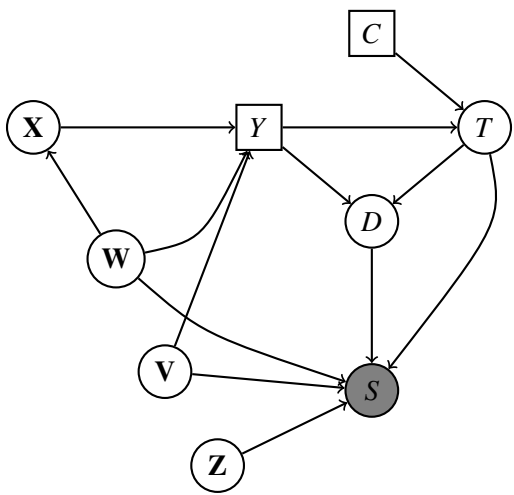

By definition, selection in the NCC study depends on all matching factors $\bM$. However, the matching factors may differ in the way they influence the outcome $Y$ and its causes. Let $\bX$ denote the $p$-vector, for some $p\geq 1$, comprising $(i)$, candidate risk factors of interest measured in the NCC study, and $(ii)$, all other causes of $Y$ that are not among the matching factors $\bM$. Without loss of generality, matching factors in $\bM$ can be categorized into three subgroups, as illustrated in Figure \ref{fig:generalDAG}: $(i)$ $\bW$ corresponding to factors that may cause both $\bX$ and $Y$, $(ii)$, $\bV$ representing factors that may cause $Y$ but not $\bX$, and $(iii)$, $\bZ$ corresponding to factors that cause neither $X$ nor $Y$. Although it may be argued that matching factors should be chosen among variables that cause both $\bX$ and $Y$, the example we present in Section \ref{sec:EPIC} illustrates a situation where some ``technical'' factors, such as fasting status at blood collection, may be assumed to be causes of neither $\bX$ nor $Y$. When conditional logistic regression is planned for the analysis of the NCC, including such matching factors can be relevant, as they help ensure that cases and their controls are comparable -- for example, with respect to fasting status. However, as it is clarified below, these factors generally do not need to be accounted for when computing weights for weighted analyses. 

From standard rules of causal DAGs, we have
\begin{equation}
\begin{aligned}\label{eq:CIgen}
        (Y, C, \bX) &\indep S\, |\, (D, T, \bW, \bV) \\
      \bX &\indep S\, |\, (D, T, \bW). \\
\end{aligned}
\end{equation}
These conditional independences correspond to statements of conditional ignorability of selection  \cite{rabe-hesketh_ignoring_2023}, which are valuable to determine which matching factors must be accounted for, and which ones can be ignored, when computing weights, depending on the target estimand. The following paragraphs provide details on the estimands considered in our work. We start by a detailed proof of Equation \eqref{eq:GenPrincipleWeightedAnalyses}, which is instrumental in the subsequent paragraphs.

\subsection{Proof of \eqref{eq:GenPrincipleWeightedAnalyses}}
We prove the result stated in Equation \eqref{eq:GenPrincipleWeightedAnalyses}, replacing $\bM$ with $\bM_0 = (\bW, \bV)$. The result stated in Equation \eqref{eq:GenPrincipleWeightedAnalyses} follows from similar arguments after observing that, under the DAG presented in Figure \ref{fig:generalDAG}, we also have $(Y, C, \bX) \indep S\, |\, (D, T, \bM)$, with $\bM = (\bW, \bV, \bZ)$.  
By successively applying the law of total expectation and the first conditional independence statement in Equation \eqref{eq:CIgen}, and assuming that $0<\P(S=1 | D, T, \bM_0)<1$ almost surely, it follows that, for any integrable function $\phi$, we have
\begin{align*} 
\hspace{-25pt}&\E \bigg[ \frac{\phi(\bX, Y)\1(S=1)}{\P(S=1 | D, T, \bM_0)} \bigg] \\
&\quad\quad\quad = \E \bigg\{ \E\bigg[\frac{\phi(\bX, Y) \1(S=1)}{\P(S=1 | D, T, \bM_0)}  \Big| D, T, \bM_0 \bigg] \bigg\} \\
&\quad\quad\quad = \E \bigg\{  \frac{\E[\phi(\bX, Y) |  D, T, \bM_0] \E[\1(S=1)|  D, T, \bM_0]}{\P(S=1 | D, T, \bM_0)}\bigg\} \\
&\quad\quad\quad = \E[\phi(\bX, Y)],
\end{align*}
which is the result stated in Equation \eqref{eq:GenPrincipleWeightedAnalyses}, with $\bM_0 = (\bW, \bV)$ instead of $\bM$.

\subsection{Estimation of the association between covariates} \label{sec:WeightsCorrTheo}
For simplicity, we consider the estimation of the linear regression of one of the covariates, $X_k$ for some $k\in \{1, \ldots, p\}$, onto other covariates $\bX_\ell$, with $\ell \subseteq \{1, \ldots, p\} \setminus \{k\}$. We assume that $\mathbb{E}[||\bX_\ell||^2] < \infty$ and $\mathbb{E}[X_k^2] < \infty$, and that the matrix \( \mathbb{E}[\bX_\ell^\top \bX_\ell] \) is positive definite. We aim to estimate the population linear regression coefficient
\[
\theta^*= \argmin_{\theta \in \mathbb{R}^p} \mathbb{E}[(X_k - \bX_\ell^\top \theta)^2].
\]

We define the inverse probability weighted estimator
\[
\hat{\theta} = \left( \sum_{i=1}^n \tilde w_i \bX_{\ell,i} \bX_{\ell,i}^\top \right)^{-1}
\left( \sum_{i=1}^n \tilde w_i \bX_{\ell,i} X_{k,i} \right),
\]
where we use the notation $\tilde w_i  = S_i/\P(S=1 | D= D_i, T= T_i, \bW=\bW_i)$. We recall that $n$ denotes the size of the originating cohort, and that $S_i=1$ indicates selection into the NCC.

Assuming that $0<\P(S=1 | D, T, \bW)<1$ almost surely, similar arguments to those above yield
\begin{align*} 
\hspace{-25pt}&\E \bigg[ \frac{\psi(\bX)\1(S=1)}{\P(S=1 | D, T, \bW)} \bigg] = \E[\psi(\bX)], 
\end{align*}
for any integrable function $\psi$. Here, we made use of the second conditional independence statement in Equation \eqref{eq:CIgen}. In particular, 
\[
\mathbb{E} \left[ \frac{S}{\P(S=1 | D, T, \bW)} \bX_\ell X_k \right] = \mathbb{E}[\bX_\ell X_k], \quad
\mathbb{E} \left[ \frac{S}{\P(S=1 | D, T, \bW)} \bX_\ell ^\top  \bX_\ell\right] = \mathbb{E}[\bX_\ell^\top \bX_\ell].
\]

Therefore, the parameter \( \theta^* \) satisfies:
\[
\theta^* = \left( \mathbb{E}[\bX_\ell^\top\bX_\ell] \right)^{-1} \mathbb{E}[\bX_\ell X_k] = 
\left( \mathbb{E}\left[ \frac{S}{\P(S=1 | D, T, \bW)} \bX_\ell^\top\bX_\ell \right] \right)^{-1}
\mathbb{E} \left[ \frac{S}{\P(S=1 | D, T, \bW)} \bX_\ell X_k \right].
\]

From the law of large numbers and the assumptions above, it follows that
\[
\frac{1}{n} \sum_{i=1}^n \tilde w_i \bX_{\ell, i}X_{k,i} \xrightarrow{\P} \mathbb{E}[\bX_\ell X_k],
\quad
\frac{1}{n} \sum_{i=1}^n \tilde w_i\bX_{\ell, i} \bX_{\ell, i}^\top \xrightarrow{\P} \mathbb{E}[\bX_{\ell}^\top\bX_\ell ].
\]

Since \( \mathbb{E}[\bX_\ell^\top\bX_\ell] \) is invertible by assumption, we also have

\[
\left( \frac{1}{n} \sum_{i=1}^n \frac{S_i}{\P(S=1 | D_i, T_i, \bW_i)} \bX_{\ell, i} \bX_{\ell, i}^\top \right)^{-1} \xrightarrow{\P} 
\left( \mathbb{E}[\bX_\ell^\top\bX_\ell] \right)^{-1}.
\]

Therefore, by Slutsky’s theorem,
\[
\hat{\theta} \xrightarrow{\P} \theta^*.
\]

Because $D$ and $T$ still have to be accounted for in the compuation of weights $1/\P(S=1|D, T, \bW)$, controls with short follow-up typically get large weights. Consequently, those controls that are still selected in the NCC may influence the estimation of $\theta^*$ heavily. It is noteworthy that if $\bX$ is not a cause of $Y$, then $\bX \indep S\, |\, W$. In such cases, weights derived from estimates of $\P(S=1 | \bW)$ yield valid estimates of the association between components of $\bX$, and more generally, of $\E[\psi(\bX)]$ for any integrable function $\psi$, while not suffering from the limitations of weights based on $T$ and $D$. 

\subsection{Estimation of the conditional survival function}\label{sec:AppendixCondSurv}
We now focus on the estimation of the conditional survival function $\P(Y\geq t |X_k=x_0)$ for some $t\geq 0, k\in\{1, \ldots, p\}$ and $x_0\in\R$. We assume that $Y \indep C | X_k$, and, for simplicity, that $X_k$ is categorical and $p_0 = \P(X_k =x_0)>0$. The proof for the unconditional survival function follows from similar arguments and is omitted. 

Consider the following weighted conditional Kaplan-Meier estimate
$$\hat S_n(t) = \prod_{D_i=1, T_i\leq t, X_{k,i} = x_0} \Big(1- \frac{\tilde w_i}{\sum_{k:T_k \geq T_i, X_{k,i}=x_0} w_k}\Big),$$ 
with, this time,  $\tilde w_i  = S_i/\P(S=1 | D_i, T_i, \bW_i, \bV_i)$. 

Introduce, the following two weighted count processes
$$ N_n(t) = \sum_{i=1}^n \tilde w_i \1(T_i\leq t, D_i=1, X_{k,i}=x_0)$$
$$ Y_n(t) =  \sum_{i=1}^n \tilde w_i \1(T_i\geq t, X_{k,i}=x_0), $$
defined over $[0, \tau(x_0)]$ with $\tau(x_0) =  \sup\{ t\geq 0: S_T(t) =\P(T\geq t| X_k =x_0) >0 \}$. 

The two classes of functions ${\cal F}_N= \{\1\{T\leq t, D=1, X_k= x_0: t\in[0, \tau]\}\}$ and ${\cal F}_Y= \{\1\{T\geq t, X_k= x_0: t\in[0, \tau]\}\}$, are VC-classes, and because weights $\tilde w_i$ are bounded, the weighted versions of these functions are Glivenko-Cantelli \cite{vaartwellner1996}. 
Then, applying similar arguments to those above, it follows that 
$$ \sup_{t\in[0, \tau(x_0)]} \left|N_n(t) - p_0\P(T\leq t, D=1| X_k =x_0)\right| \xrightarrow{\P} 0$$
and
$$\sup_{t\in[0, \tau](x_0)} \left| Y_n(t) - p_0S_T(t) \right|\xrightarrow{\P} 0.$$
By standard arguments from counting process theory\cite{andersenetal1993}, we obtain
$$\hat S_n(t) = \exp\Big(-\int_0^t \frac{dN_n(u)}{Y_n(u)}\Big) + o_\P(1), $$
for any $t< \tau(x_0)$.
Eventually, denoting the conditional hazard rate of $Y$, given $X_k=x_0$ at time $u$ by $\lambda(u|X_k=x_0)$, we obtain
$$\hat S_n(t) \xrightarrow{\P}  \exp\Big(-\int_0^t \lambda(u|X_k=x_0)\Big)= \P(Y\leq t| X_k =x_0), $$
uniformly over $t\in[0, \tau(x_0)[$.

\subsection{Estimation of log-hazard ratios under Cox PH models}
In this paragraph, we work under the following Cox PH model 
$$ \lambda(t | \bz) =  \lambda_0(t)\exp(\bz^T \alpha^*)$$
for some variable $\bz \in \R^d$ and $\alpha^*\in\R^d$, for some $d\geq 1$.

Set $\tilde w_i  = S_i/\P(S=1 | D_i, T_i, \bW_i, \bV_i)$ as in \ref{sec:AppendixCondSurv}, $ Y_j(t) = \mathbf{1}_{\{ T_j \geq t \}}$, and let $(\bZ_1, \ldots, \bZ_n)$ represent the $n$ observations of the variable $\bz$.  
Introduce, for $k=0, 1, 2$, $$S_n^{(k)}(t, \alpha) = \sum_{j=1}^n \tilde w_j Y_j(t) \exp(\bZ_j^\top \alpha) \bZ_j^{\otimes k},$$
with $\bZ^{\otimes 0} = 1$, $\bZ^{\otimes 0} = \bZ$, and $\bZ^{\otimes 2} = \bZ\bZ^T$, 
and 
$$ \bar Z_n(t, \alpha) = \frac{S_n^{(1)}(t, \alpha) }{S_n^{(0)}(t, \alpha) }$$

Define $\hat \alpha$ as the root of the weighted partial likelihood score function,
\[
U_n(\alpha) = \frac{1}{n} \sum_{i=1}^n \tilde w_i \int_0^\tau \left[ \bZ_i -  \bar Z_n(t, \alpha) \right] dN_i(t)
\]
with $\tau =\sup\{t\geq 0: \P(T\geq t) > 0\}$.

Now consider the population versions of the quantities introduced above. For $t\in[0, \tau]$, $\alpha$ in a compact $A$ around $\alpha^*$, and $k=0, 1, 2$, define 
$$ s^{(k)}(t, \alpha) = \E\left[ Y(t) \exp(\alpha^T \bZ) \bZ^{\otimes k} \right], $$
Assuming that $s^{(0}(t, \alpha)>0$ for all $t\in[0, \tau]$ and $\alpha\in A$, further set $\bar z(t,\alpha) =s^{(1)}(t, \alpha)/s^{(0)}(t, \alpha)$, and introduce the matrix
$$ v(t, \alpha) = \frac{s^{(2)}(t, \alpha)}{s^{(0)}(t, \alpha)} - \bar z(t,\alpha)^{\otimes 2}.$$

Finally introduce \( U(\alpha) = \E\left[ \int_0^\tau (Z - \bar z(t, \alpha))dN(t)\right] \).

Observing that classes of functions $\{Y(t)\exp(\alpha^T\bZ)\bZ^{\otimes k}, t\in[0, \tau], \alpha\in A\} $ are Glivenko-Cantelli, and weights are bounded, similar arguments as above lead to
$$ \sup_{t\in[0, \tau], \alpha\in A} |S_n^{(k)}(t, \alpha) - s^{(k)}(t,\alpha)|\xrightarrow{\P} 0, \quad k=0, 1, 2.$$
Consequently, 
\[
\sum_{\alpha \in A}| U_n(\alpha) - U(\alpha) | \xrightarrow{\P} 0, 
\]
and then, further assuming that $ I(\alpha) = \int_0^\tau s^{(0)}(t,\alpha) v(t, \alpha)\lambda_0(t)dt$ is positive definite at $\alpha^*$, (and, therefore, $\alpha^*$ is the unique zero of $U$), we eventually have $\hat{\alpha} \xrightarrow{\P} \alpha^*$ by standard consistency of Z-estimators.

\end{document}